\documentclass{elsart}

 \usepackage{epsfig}

\usepackage{amssymb}
\begin{document}
\begin{frontmatter}


\title{  Glassy dynamics    in the HMF model  }


\author
{Alessandro Pluchino*},   \corauth[cor1]{Corresponding author: alessandro.pluchino@ct.infn.it}
\author
{Vito Latora},   
\author
{Andrea Rapisarda} 

\address{Dipartimento di Fisica e Astronomia,  Universit\'a di Catania,\\
and INFN sezione di Catania, Via S. Sofia 64,  I-95123 Catania, Italy}

\begin{abstract}


We discuss the glassy dynamics recently found in the
meta-equilibrium quasi stationary states (QSS) of the HMF model.
The relevance of the initial conditions and the connection with
Tsallis nonextensive thermostatistics is also addressed.

\end{abstract}

\end{frontmatter}

\section{Introduction}

\label{intro}

In this paper we present a brief review of the glassy and
anomalous behavior observed in the dynamics of the Hamiltonian
Mean Field (HMF) model \cite{plu1,plu2}: a simple XY model of
fully-coupled inertial spins with ferromagnetic long-range
interactions \cite{ant1,lat1,lat2,lat4,leshouches,celia1}. We
show, in particular, a more detailed description of the
microscopical analogies between the quasi-stationary states (QSS)
regime found in the HMF model and the spin glass phase scenario of
the Sherrington-Kirkpatrick (SK) infinite-range model
\cite{sk1,sk2}. We also discuss the importance of the initial
conditions in order to
observe {\it dynamical frustration} \cite{plu2}.  
The latter is a crucial feature for the emergence of a glassy
dynamics,  since, a-priori, the  HMF model  is not frustrated.
 Dynamical
frustration is related to the weak-ergodicity breaking phenomenon,
typical of glassy-systems, \cite{bou0,bou} and to other dynamical
anomalies, such as superdiffusion and L\'evy walks, negative specific heat,
 vanishing
Lyapunov exponents, non-Gaussian velocity pdfs, power-law decaying
correlation functions\cite{plu2,lat4,leshouches,celia1}.
This anomalous behavior seems to be linked
to the fractal structure of the region of phase space
 in which the
systems remains trapped when the dynamics starts sufficiently far
from equilibrium.
We will show that such a dynamics can be quantitatively
characterized by the introduction of  a new order parameter,
namely the {\it polarization $p$} \cite{plu1}.
In the end we will also briefly discuss the links with
Tsallis nonextensive thermostatistics
scenario\cite{tsa1,tsa2,cho}.

\section{Glassy phase and nonextensivity in the HMF model}
\subsection{The model}
The HMF model, here considered in its ferromagnetic version,
consists  of N planar classical spins
${\stackrel{\vector(1,0){8}}{s_i}}=(cos\theta_i, sin\theta_i)$
interacting through an infinite-range potential \cite{ant1}. The
Hamiltonian is
\begin{equation}
\label{eq1}
        H= K+V
= \sum_{i=1}^N  {{p_i}^2 \over 2} +
  {1\over{2N}} \sum_{i,j=1}^N  [1-cos(\theta_i -\theta_j)]~~,
\end{equation}
\noindent
 where $-{\pi}<{\theta_i}<{\pi}$ is the  angle of the  $ith$ spin and $p_i$ the
conjugate variable representing the rotational velocity. Since the
modulus of each spin is unitary, we can represent the system of N
planar rotating spins  as N interacting particles moving on the
unit circle. The usual order parameter of the model is the
magnetization M:
\begin{equation}
\label{eq2} M = {1\over{N}} | \sum_{i=1}^N
\stackrel{\vector(1,0){8}}{s_i} |~~~~~.
\end{equation}
%
The equilibrium solution in the canonical ensemble predicts a
second-order phase transition from a low-energy condensed
(ferromagnetic) phase with magnetization  $M\ne0$, to a
high-energy one (paramagnetic), where the spins are homogeneously
oriented on the unit circle and $M=0$. The {\em caloric curve},
i.e. the dependence of the energy density $U = E/N$ on the
temperature $T$, is given by
$U = {T \over 2} + {1\over 2} \left( 1 - M^2 \right)
~$\cite{ant1,lat1}.
The critical point is at energy density $U_c=\frac{3}{4}$,
corresponding to a critical temperature $T_c=\frac{1}{2}$
\cite{ant1}.

The dynamics of HMF shows several anomalies before complete
equilibration.
More precisely,
if we adopt the so-called $M1$ initial
conditions, i.e. $\theta_i=0$  for all $i$ ($M(0)=1$) and
velocities uniformly distributed ({\it water bag}), the results of
the simulations, in a special region of energy values ($\frac{1}{2}
<U<U_c$), show a disagreement with the canonical prediction for a
transient regime whose length depends on the system size N. In
such a regime, the system remains trapped in metastable states
(QSS) at a temperature lower then the canonical equilibrium one,
until it slowly relaxes
 towards Boltzmann-Gibbs (BG)  equilibrium, showing strong memory effects.
This transient regime becomes stable if one takes first the infinite size limit
and then the infinite time limit
\cite{lat4}.

  \subsection{Glassy dynamics}

The observation of these long relaxation times and in particular
of aging \cite{plu2,celia1} for the QSS  was the first indication
towards  a possible interpretation of this regime in terms of
glassy dynamics . The paradigmatic example of this behavior  are
spin glasses \cite{bou,bert}. In the materials that originally
were called {\it 'spin glasses'} the randomly distributed magnetic
impurities determine a random distribution ('quenched disorder')
of ferromagnetic and anti-ferromagnetic interactions among the
magnetic spins, thus generating frustration in the lattice.
In these systems the impossibility
to minimize simultaneously the interaction energies of all the
couple of spins leads to a frustrated situation, which determines
a very complex energetic landscape in phase space. The latter
appears as consisting of large valleys separated by high
activation energies. Each valley contains many local minima in
which the system, at low temperature, can remain trapped for a
very long time. This time grows exponentially with the height of
the energy barriers, thus the system shows very  slow relaxation,
strong memory effects and aging.
In an ordinary ferromagnetic phase, where there is only one energy
minimum, the application of an external magnetic field gives
suddenly rise to a non-zero magnetization. The latter, for a fixed
temperature, remains constant until the field is active and then
vanishes very rapidly. On the contrary, in the spin glass phase
the magnetization shows a strong dependence on the thermal history
of the system (aging). After quenching the spin glass below its
critical temperature in presence of the external field, the system
settles in at a particular magnetization value ({\it field cooled
magnetization})
 that does not change instantaneously when the field is switched
 off, but relaxes to equilibrium very slowly. This relaxation
 depends on the waiting time spent between the quenching and the
 elimination of the external field. Such a behavior can be
explained within the so-called {\it weak-ergodicity breaking} framework
 \cite{bou0,bou}.
A very similar situation seems to happen in the QSS regime of the
HMF model\cite{plu1,plu2}.
%
%
%
\begin{figure}
\label{fig1}
\begin{center}
\epsfig{figure=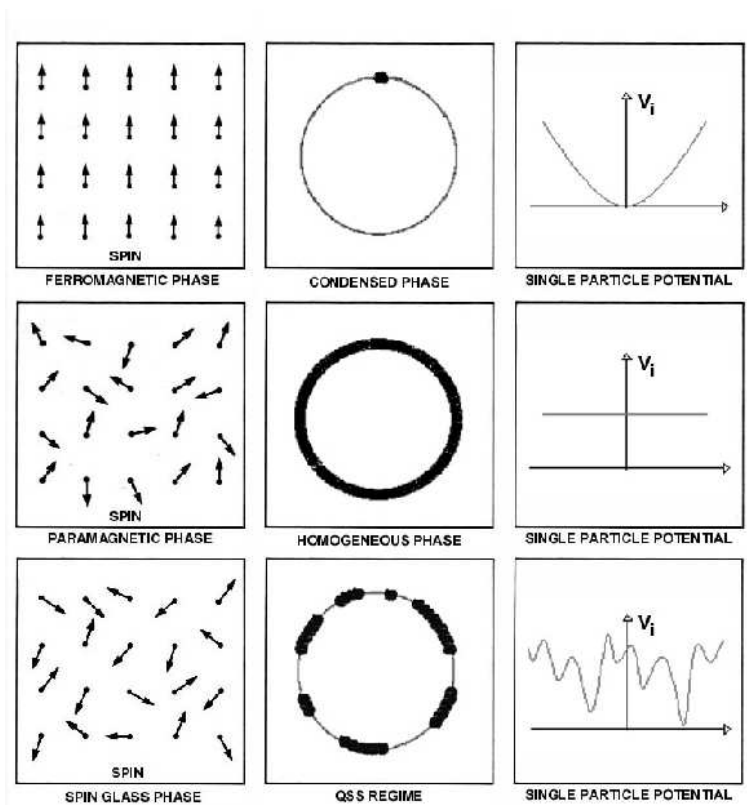,width=10truecm,angle=0}
\end{center}
\caption{The figure shows a schematic representation of the three
phases of a spin glass system: ferromagnetic phase, paramagnetic
phase and spin glass  phase. In the first column spins are
represented in a two-dimensional lattice. In the second column, in
analogy with the HMF model, spins are represented as particles
rotating on the unit circle.
 In the third column we draw the corresponding schematic single-particle
 potential landscape of
 the three  phases.}
\end{figure}
\begin{figure}
\label{fig2}
\begin{center}
\epsfig{figure=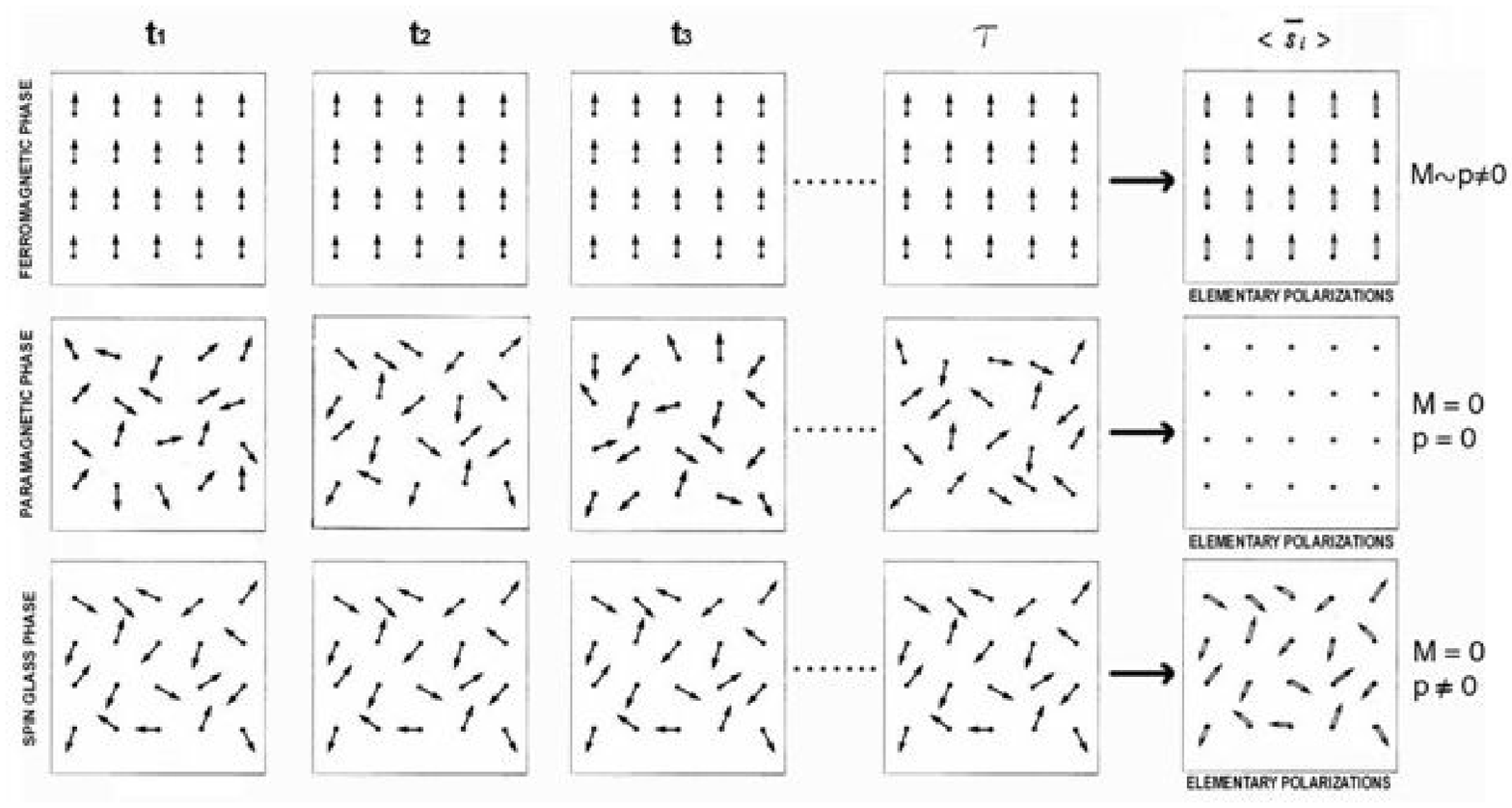,width=14truecm,angle=0}
\end{center}
\caption{In this figure we show a temporal sequence of snapshots
for each of the three phases of a spin glass. Only comparing the
different
 snapshots in the sequences it is possible to distinguish the
 paramagnetic phase (where
 the snapshots change in time),  from the spin glass one (where all
 the snapshots are identical).
 In the last column we report the elementary polarizations resulting
 for each phase.
  By averaging their modulus over all the spins of the lattice we obtain
  the  order
  parameter $p$, see text, which allows  to discriminate between the three phases.}
\end{figure}
%
 Within the mean-field
framework of the {\it Sherrington-Kirkpatrick} (SK) model
\cite{sk1,sk2}, the first solvable model of a spin glass system with
gaussian distribution of interactions, it was possible to observe
three different phases, namely, paramagnetic (PA), ferromagnetic
(FE) and spin glass (SG) phase, depending on the temperature and
the parameters of the Gaussian distribution. Each phase is
characterized by a different microscopic behavior and a different
kind  of orientation order.
Although today some features of the SK model are considered rather
obsolete,  its microscopic interpretation of the SG phase can be
still considered  as representative of a generic glassy-like
phase. Thus, in order to get an intuitive picture of the
differences between the three phases, let us consider for example
a two-dimensional lattice of  planar spins,  see first column of
fig.1.
This schematic picture describes also the  HMF dynamics if one imagines to
locate the spins in a square lattice.
%
%
Now let us  take some snapshots of the spin configuration in
each of the three phases, see fig.2. In the FE phase ($T<T_c$) all
the spins results aligned and frozen in their equilibrium
position, so it is easy to recognize this phase even by means of
snapshots taken for only one particular instant of time. But in
this   way it would be impossible to distinguish between the PA
and the SG phase. In fact in both these phases the instantaneous
mutual orientations of the spins are random, in the PA phase
($T>T_c$) due to the high temperature and in the SG phase
($T<T_c$) due to the quenched disorder of the interactions. So we
necessarily need to consider a temporal sequence of snapshots in
order to discriminate the SG from the PA phase. In the SG phase
all the snapshots will be identical with each other, since each
spin is frozen and retains the same orientation over very long
periods of time.
On the other hand, in the PA phase the orientation of {\it the
same spin} at successive instants of time changes randomly.
 It appears clearly that
the magnetization order parameter, calculated as in eq.(2) at one
instant of time, vanishes in the SG phase just like in the PA one.
Therefore, in order to discriminate between spin glass disorder
and paramagnetism, one needs an additional order parameter. The
latter   should take into account the temporal evolution of each
spin, in order to measure its degree of freezing.
In effect  a parameter of this kind, called {\it 'EA order
parameter'}, was originally proposed in refs. \cite{sk1,sk2}, although
later it turned out to be inadequate for the mean-field
theoretical description of the SG phase \cite{bou}.
Nevertheless, inspired by the physical meaning of this parameter,
we have proposed  a new order parameter  in the context of the HMF
model, the {\it polarization $p$}  to characterize
in a quantitative way the 
glassy dynamics of  the QSS regime\cite{plu1}.
\subsection{The Polarization}
We define the {\it elementary polarization} as the temporal
average, integrated over an opportune time interval $\tau$, of the
successive positions of each elementary spin vector: 
\begin{equation}
\label{eq3}
<\stackrel{\vector(1,0){8}}{s_i}>={1\over{\tau}} \int_{t_0}^{t_0+\tau}
\stackrel{\vector(1,0){8}}{s_i}(t)dt~~~~~~i=1,...,N ~~,
\end{equation}
being $t_0$ the  initial transient time.
Then we further average the modulus of the elementary polarization
over the N spin configuration, to finally obtain the {\it average
polarization p}:
\begin{equation}
\label{eq4}
{\it p}={1\over{N}} \sum_{i=1}^N  | <\stackrel{\vector(1,0){8}}{s_i}>|~~~~.
\end{equation}
%
\begin{figure}
\begin{center}
\label{fig3} \epsfig{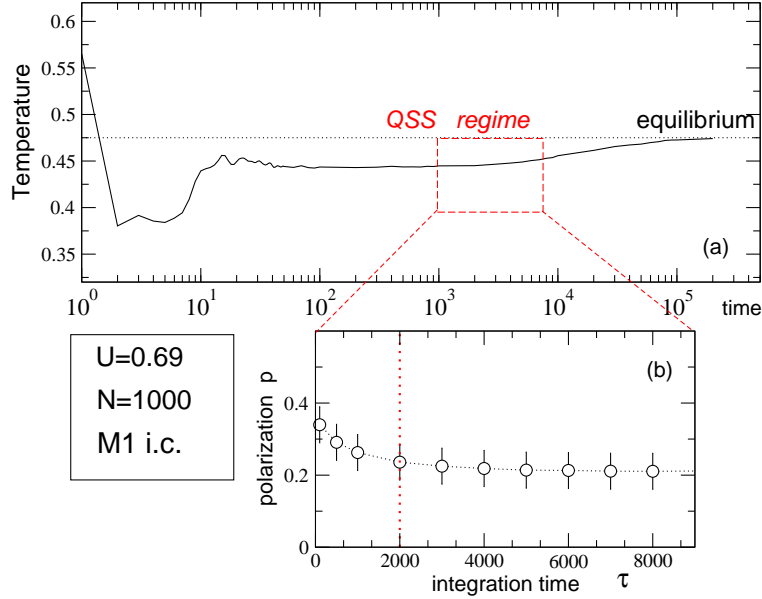}
\end{center}
\caption{In the upper panel of the figure we plot the temporal
evolution of temperature for the HMF model for U=0.69, N=1000 and
M1 initial conditions. In the lower panel we show the value of
polarization versus the integration time $\tau$ (in a linear
scale), after a transient time $t_0= 1000$ time units and for a window of
9000 time units. One can see that  for $\tau$ greater than 2000 -
i.e. the standard interval we use for our simulations - the
polarization does not change significatively up to the end of the
QSS temperature plateau. The values of polarization were averaged
over 20 different realizations - the error bars refers to such an
average.}
\end{figure}
%
%
It is easy to see (last column of fig.2) that:
\begin{enumerate}
 \item in a pure ferromagnetic (condensed) phase each elementary
polarization vector coincides with the correspondent spin vector,
both being frozen and parallel, then the average polarization {\it
p} keeps a non zero value equal to the modulus of the average
magnetization per spin M;
\item in a paramagnetic (homogeneous) phase, where the orientation
of each spin vector at every time changes in a completely aleatory
way, this continuous motion yields a vanishing value for both M
and the average polarization;
\item in a spin glass phase, where the spatial disorder is random
but the dynamics is quenched, while $M$ vanishes as in the PA
phase, {\it p} gets a non zero value as in the FE one.
\end{enumerate}
%
%
%
\begin{figure}
\begin{center}
\epsfig{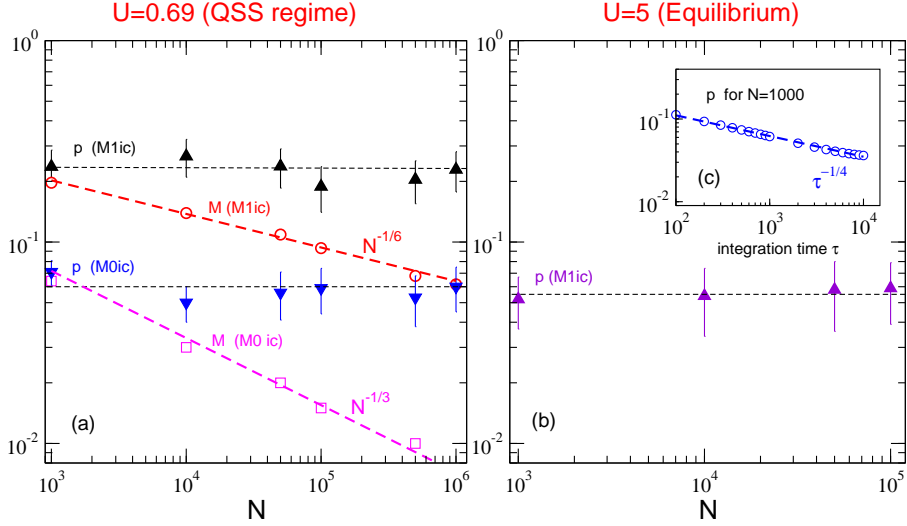}
\end{center}
\caption{The figure shows the behavior  of $p$ and $M$ with the
size  of the system. On the left (a) we plot the scaling in the
QSS regime at $U=0.69$ for the two different initial conditions
considered in the paper, $M1$ and $M0$. While the magnetization
tends to zero going towards the infinite size limit, the
polarization remains constant. The polarization $p$ is
significatively different from zero only for $M1$ initial
conditions, see text for further details. On the right (b), the
behavior of p is plotted only for $M1$ initial conditions at the
overcritical energy $U=5$, i.e. in the full paramagnetic
(homogeneous) phase, where the system reaches immediately the BG
equilibrium. In this case the polarization is very small and
almost equal to the  case $U=0.69$ with $M0$
initial conditions. Moreover, in this case, as shown in the inset,
at variance with the behavior plotted  in fig.3 for the QSS
regime, the polarization vanishes,  as $\tau^{-1/4}$ for N=1000,
increasing the integration time interval $\tau$.}
\end{figure}
%
%
From the numerical simulations, it results that the QSS temperture
lies on the extension of the high-temperature line of the caloric
curve below $T_c$  \cite{plu2,lat4}. This implies that in the QSS
regime $M$ vanishes with the size $N$ of the system (more
precisely as ${N}^{-1/6}$), so we have ${M\simeq 0}$ below the
critical temperature, just as in the SG phase of the SK
model\cite{sk1,sk2}. Thus, the next natural step is to check if the
polarization order parameter would remain different from zero in
the QSS regime for a growing size of the system.
Preliminarly we consider   the
calculation of $p$  versus the integration time interval
$\tau$ at $U=0.69$ and $N=1000$, after a transient time $t_0= 1000$.
As one can see in fig.3,
lower panel, the value of the polarization does not change
significatively increasing the integration time interval $\tau$  beyond
$\tau=2000$,  up to the
end of the QSS temperature plateau, see upper panel.  The same behavior is
obtained for greater values of $N$.
In the following,   we adopt the time interval $\tau=2000$ for the
the calculation of $p$.
Starting the numerical simulations from the usual M1 initial
conditions,  we have  found ( see the upper part of fig.4(a) )
 that, in the QSS regime, while M goes to zero with the expected scaling,
 the polarization
 $p$ does not vanish and remains constant inside the error: $p\sim0.24$.
 This finite value of $p$ which characterizes quantitatively a frozen dynamics,
  is due to a 'dynamical frustration' phenomenon \cite{plu1}:
  in fact the QSS are characterized by the
  presence of many clusters of particles
 appearing and disappearing on the unit circle, see the lower picture
  in the second column of
 fig.1. Each of them compete with the others trying
   to capture as many particles as possible
 in order to relax to the equilibrium configuration with a magnetization
 $M \sim 0.3$.
These results are also in perfect agreement with the observed
dynamical correlations in the $\mu$-space \cite{plu2,lat4}:  as
required by the weak-ergodicity breaking hypotesis, during the QSS
regime the
 system lives in a smooth fractal part of the {\it a-priori} accessible
 phase space \cite{lat4},
  and for $N$ going to infinity it never escapes from that region. So, in the
thermodynamic limit,
 the QSS regime can be considered as a new
 {\em glassy phase} of  the HMF model.
As expected, when the dynamical frustration disappears, i.e. when
the system (for $N $ finite) has reached the equilibrium
conditions of the condensed phase, we loose any trace of
glassy-like dynamics and one obtains values
of $M$ and $p$ which are equal everywhere but  not zero \cite{plu1}. Finally,
in the full homogeneous phase both $M$ and $p$ vanish, because the
spins can rotate freely \cite{plu1}.

\subsection{The role of initial conditions}
It is important to stress the role of the M1 initial condition in
order to have weak-ergodicity breaking and glassy behavior. In
fact, if we start from initial conditions with both angles and
velocities uniformly distributed (namely M0 initial conditions,
since M(t=0)=0), the QSS regime  shows a very different scenario:
in fact in this case   neither power-law correlation functions nor
dynamical structures in the $\mu$-space are present \cite{plu2}.
 Such a scenario is consistent with the different value  of the
polarization calculated in such QSS regime reached from M0 initial
conditions, see lower part of fig.4 (a). One can see that in this
case the values of $p$ vs. $N$ is constant to a value  much
smaller than before, i.e. $p\sim6\cdot10^{-2}$. This is also the
order of magnitude of the polarization at  equilibrium in the full
homogeneous phase (for M1 initial conditions), see fig.4 (b).
Please  note also that here, for fixed $N$ $(1000)$, the value of
the polarization vanishes with the integration
 time interval $\tau$ as $\tau^{-1/4}$,  see the inset.
The intuitive explanation of such a different behavior is quite
simple. Starting from $M0$ initial conditions, although we
 are far from Boltzmann Gibbs (BG) equilibrium, we do not have the same kind
 of {\it kinetic explosion}, as
 for M1 initial conditions, which creates
 the long-lasting dynamical correlations.
 In fact, in this case
the system is directly put on the QSS plateau at a temperature
 $T= 0.38$  where M(0)=0 and thus also
  the force acting on each spin,  proportional to $ M $
 \cite{lat4}, vanishes  since the beginning.
For M0 initial conditions  we do not
  have  any kind of fast quenching
 from an high temperature phase, at variance with  the M1 case,
 and therefore we do not find  any glassy-like
 behavior, dynamical frustration or weak-ergodicity breaking.
On the other hand several other dynamical anomalies observed in
connection with the M1 case (fractal-like structures in the
$\mu$-space, power law velocity pdfs and correlation functions,
L\'evy walks and superdiffusion, aging) have not been found for
the M0 one\cite{plu2}. This suggests that a connection with
Tsallis nonextensive thermodynamics \cite{lat4,plu2}, exists
probably only for the QSS regime obtained starting the system with
M1 and not with M0 initial conditions. The metastable states
 in this second case (M0) have a different microscopic nature and
 can be probably better interpreted as
 Vlasov stationary states
\cite{thierry}.

\subsection{Links to Nonextensive Thermostatistics  }

In ref.\cite{lat4} we had already found  a link of the QSS regime
with Tsallis nonextensive thermostatistics, by reproducing  the
microcanonical non-Gaussian  velocity  pdfs with a q-exponential
curve. However  the value of q obtained in that case is rather
large and not fully understood. A very  interesting progress in
that direction has been presented, considering the more
appropriate canonical ensemble, by Baldovin\cite{baldo}. On the
other hand,
 we have recently found that also the power-law decay of
correlation functions, from the QSS regime to equilibrium, can be
very well explained by q-exponential curves\cite{plu2,sugi}. More
interesting is the fact that in this case, we obtain
$q=1.60\pm0.05 $ for  the entropic index. In fact this is the
value   expected from the relationship, derived in ref.
\cite{tsabuk}, between $q$ and the anomalous diffusion exponent
$\alpha$, i.e.  $~~q=\frac{3 \alpha -2}{\alpha}$. In our case we
had previously found a value $\alpha = 1.4\pm 0.2$ for
superdiffusion in the the QSS regime \cite{lat2}, thus in this
respect the nonextensive formalism seems to apply in a consistent
way. A more detailed study in this direction is in progress.
%

\section{Conclusions}

In this paper we have shown that the  metastable quasi-stationary
states of the HMF model, obtained from M1 initial conditions, can
be interpreted as a glassy phase of the system. This phase can be
characterized by a new order parameter, the polarization p, which
gives a quantitative description of the frozen dynamics. This fact
establishes a very interesting and promising relationship between
nonextensive systems  and glassy ones,  which  will hopefully
 lead to new exciting discoveries in the near future.

\vskip 0.2truecm \noindent We thank F. Baldovin, C. Tsallis and M.
M\'ezard for stimulating discussions.

\bigskip

\noindent


\begin{thebibliography}{00}




\bibitem{plu1}  A. Pluchino,  V. Latora,  A. Rapisarda,
[cond-mat/0306374], submitted.

\bibitem{plu2}  A. Pluchino,  V. Latora,  A. Rapisarda,
[cond-mat/0303081], Physica D in press.

\bibitem{ant1} M. Antoni  and S.Ruffo, Phys.  Rev. E {\bf 52} (1995) 2361.


\bibitem{lat1} V. Latora , A. Rapisarda  and S. Ruffo,
Phys. Rev. Lett. {\bf 80}  (1998) 692 and  Physica D {\bf 280} (1999) 81. 


\bibitem{lat2} V. Latora, A. Rapisarda  and S. Ruffo,
  Phys. Rev. Lett. {\bf 83}  (1999) 2104 


\bibitem{lat4} V. Latora , A. Rapisarda  and C. Tsallis, Phys. Rev.
E   {\bf 64}  (2001)  056134 .

\bibitem{leshouches} {\it Dynamics and Thermodynamics of Systems with Long Range
interactions} T. Dauxois, S. Ruffo, E.  Arimondo, M. Wilkens
Eds., Lecture Notes in Physics Vol. 602, Spinger (2002)


\bibitem{celia1}  M.A Montemurro,  F.A. Tamarit and C. Anteneodo,
 Phys. Rev. E  {\bf 67}  (2003) 031106.

\bibitem{sk1}  D. Sherrington and S. Kirkpatrick,   Phys. Rev. Lett. {\bf 35},
(1975) 1792
\bibitem{sk2} D. Sherrington and   S. Kirkpatrick,
  Phys. Rev. B {\bf 17} (1978) 4384.

\bibitem{bou0} J.P. Bouchaud, J. Phys. I France {\bf 2} (1992) 1705.

\bibitem{bou} J.P. Bouchaud,  L.F. Cugliandolo,     J. Kurchan,  M. Mez\'ard,
 A.P. Young ed., World Scientific Singapore   (1998).


\bibitem{bert} L. Berthier and A.P. Young, [cond-mat/0312327].


\bibitem{tsa1} G. Kaniadakis
M. Lissia and A Rapisarda Eds., proceedings of the conference
NEXT2001, special issue of Physica A {\bf 305}
(2002).


\bibitem{tsa2} {\it Nonextensive Entropy: interdisciplinar ideas},
C. Tsallis and M. Gell-Mann Eds. Oxford University press(2004)

\bibitem{cho}
 A. Cho, Science {\bf 297} (2002) 1268; Letters to the
editor S. Abe, , A.K. Rajagopal,
 A. Plastino,
 V. Latora,   A. Rapisarda, A. Robledo,  Science {\bf 300} (2003) 249.









\bibitem{thierry} T. Dauxois,  talk presented at this conference.



\bibitem{baldo} F. Baldovin,  talk presented at this conference.

\bibitem{sugi}  A. Pluchino,  V. Latora,  A. Rapisarda,
Continuum dynamics and Thermodynamics  in press.


\bibitem{tsabuk} C. Tsallis,    D.J. Bukman,
Phys. Rev. E {\bf 54}    (1996) R2197.





\end{thebibliography}
\end{document}